\documentclass[aps,showpacs,preprint,graphics]{revtex4}
\usepackage{amssymb}
\usepackage[dvips]{graphicx}
\usepackage{amsmath}
\usepackage{bm}
\usepackage{epsfig}

\begin{document}

\title{Aspects of $(4+4)$-Kaluza-Klein type theory}
\author{ $^{1,2}$ Juan Antonio Nieto \thanks{%
E-mail:nieto@uas.edu.mx } and $^{2}$ Jos\'e Edgar Madriz Aguilar\thanks{%
E-mail address: madriz@mdp.edu.ar} }
\affiliation{$^{1}$ Facultad de Ciencias F\'{\i}sico-Matem\'aticas de la Universidad
Aut\'onoma de\\
Sinaloa 80010, Culiac\'an Sinaloa, M\'exico \\
and\\
$^{2}$ Departamento de Matem\'aticas, Centro Universitario de Ciencias
Exactas e ingenier\'{\i}as (CUCEI), Universidad de Guadalajara (UdG), Av.
Revoluci\'on 1500 S.R. 44430, Guadalajara, Jalisco, M\'exico. \\
E-mail: nieto@uas.edu.mx, madriz@mdp.edu.ar }

\begin{abstract}
We develop a type of Kaluza-Klein formalism in $(4+4)$-dimensions. In the
framework of this formalism we obtain a new kind of Schwarzschild metric
solutions that via Kruskal-Szequeres can be interpreted as mirror black and
white holes. We found that this new type of mirror black and white holes
solutions in $(3+1)$-dimensions support the idea that the original
space-time can be extended to $(4+4)$-signature. Using octonions, we also discus linearized gravity in $(4+4)$-dimensions. 
\end{abstract}

\maketitle


\vskip .5cm

\section{Introduction}

There has already been considered Kaluza-Klein type theories for non-compact
spacetime \cite{k1,k2,k3} with interesting results in inflationary and
cosmological models \cite{k4,k5,k6,k7,k8,k9,k10,k11,k12,k13,k14}. In this
article we are interested in deriving a Kaluza-Klein type theory for a
spacetime of four time and four space dimensions ($(4+4)$-dimensions). Our
formalism is based on the analogue of the approach of Refs \cite%
{k15,k16,k17,k18,k19}. \newline
There are at least three physical motivations for considering $(4+4)$%
-dimensions emerging from different sources. First, considering the
splitting of the form (4 + 4) = (3 + 1) + (1 + 3) one may ask if an electron
lives in the (1+3)-world one wonders one could be the corresponding mirro
electron in (3 + 1)-dimensions. In Ref. \cite{k20} it was shown that
massless Majorana-Wely spinor satisfying the Dirac equation in $(4+4)$%
-dimensions can be identified with a massive complex spinor in $(1+3)$%
-dimensions. This means that a Majorana-Weyl massless fermion in $(4+4)$%
-dimensions can be seen as an electron in $(1+3)$-dimensions. Second, it is
interesting that in $(4+4)$-dimensions one may consider the chain of maximal
embeddings and branches, 
\begin{equation*}
so(4,4)\supset s(2,R)\oplus so(2,3)\supset so(1,1)\oplus sl(2,R)\oplus
sl(2,2).
\end{equation*}
Finally, if one makes the question why at the macroscopic level our world is 
$(1+3)$-dimensional? Surprisingly until now there is not a satisfactory
answer. Any answer needs to explain the asymmetry between time and space in
the $(1+3)$-world. In this sense it seems to us interesting to start with
the more symmetric $(4+4)$-world model and to explore its gravitational
consequences. \newline

On the other hand, it is well known that the Kruskal--Szekeres coordinates
give an alternative description of the event horizon of black-holes \cite%
{01,02}. Traditionally, one starts with the Schwarzschild metric described
by the coordinates $t$, $r$ and the angular coordinates $\theta $ and $\phi $
associated with a unit sphere and after several algebraic steps one computes
the Kruskal--Szekeres coordinates $T$ and $X$ which become functions of $t$
and $r$. The key result of this process is that while the Schwarzschild
metric is singular at the horizon $r=2GM=r_{s}$ the Kruskal--Szekeres metric
is not. However, the final transformations between the coordinates $t$, $r$
and $T$ and $X$ seems to be, in a sense, intriguing because instead of
describing only two regions ($2$-region) as in the case of Schwarzschild
black-holes (interior and exterior) one ends up with four regions ($4$%
-region) in the Kruskal--Szekeres black-hole (see Refs. \cite{03} and \cite%
{04} for details).

In an effort of having a better understanding of the above intriguing result
of $2$-region and $4$-region we discover that one may start with a more
general Kruskal--Szekeres transformation between $t$, $r$ and $T$, $X$ which
turns out that describe no only previous $4$-region but surprisingly a total
of $8$-region. This, of course, it may be interpreted as an extra
complication. However, in this work we show that such a $8$-region
transformation suggests that the $(1+3)$-signature of the original
space-time can be extended to $(4+4)$-signature. It turns out that the $%
(4+4) $-signature can be splitted as $(4+4)=(1+3)+(3+1) $, given the
original $(1+3)$-signature and a some kind of mirror $(3+1)$-signature. But
it has been a subject of much interest to consider invariant theories of
reversal signatures transfomation such as $(1+3)\leftrightarrows (3+1)$ \cite%
{05}. Moreover, one may ask: assuming that one has a black-hole in $(1+3)$%
-dimensions what could be the corresponding mirror black hole in $(3+1) $%
-dimensions? Thus, we argue that our extended Kruskal--Szekeres coordinates
suggest that instead of thinking the $(4+4)$-world as an exotic space-time
it may considered as a more interesting scenario in which one has not only
the ordinary black-holes in $(1+3)$-dimensions but a new type of
mirror-black-hole in $(3+1)$-dimensions.\newline

It is worth mentioning that using a reversal signature a relation between $%
(3+1)$ and $(1+3)$ signatures has already been investigated in the context
of string theory \cite{05,06} . Secondly, in $(4+4)$-dimensions there exist
Majorana-Weyl spinors \cite{07,08,09,10,11,12}. Another source of physical
interest emerges from the fact that the $(4+4)$-dimensional theory may be
obtained from dimensional reduction of a $(5+5)$-dimensional theory which
originates from the so-called $M{\acute{}} $-theory \cite{13,14} (see also
Refs. \cite{15,16,17,18,19}) which is defined in $(5+6)$-dimensions. In
fact, upon space-like compactification the $(5+6)$-dimensional theory leads
to one Type II $A%
{\acute{}}%
$ and two Type II $B%
{\acute{}}%
$ string theories which "live" in $(5+5)$-dimensions \cite{18}. Further, we
believe that our work may be useful in several physical scenarios. In
par\-ti\-cu\-lar, since the $(4+4)$-world contains the same number of times
and space coordinates one has a more symmetric scenario for considering
quantum and topological frameworks.\newline

Technically our work is organized as follows. In section I we give a brief
introduction. In section II we develop a new kind of Kaluza-Klein theory in $%
(4+4)$-dimensions. In section III, we propose the extended Kruskal-Szekeres
coordinates. In section IV we interpret the new solutions as mirror black
and white holes, thus making clear the necessity of a (4+4)-space-time
structure to have a unified description of both kind of solutions, within
the same single gravitational framework. In section V, as another non-trivial application
of our formalism, we describe a linealized gravity in (4+4)-dimensions. Finally in section VI, we give some final remarks.

\section{The Kaluza-Klein type formalism}

As it is usually done in the light of Kaluza-Klein theories, our starting
point is the metric ansatz 
\begin{equation}
\gamma _{AB}(x^{C})=\left( 
\begin{array}{cc}
g_{\mu \nu }+A_{\mu }^{i}A_{\nu }^{j}g_{ij} & A_{\mu }^{j}g_{ij}+g_{\mu \nu
}B_{i}^{\nu } \\ 
A_{\nu }^{k}g_{kj}+g_{\nu \mu }B_{j}^{\mu } & g_{ij}+B_{i}^{\mu }B_{j}^{\nu
}g_{\mu \nu }%
\end{array}
\right) ,  \label{kk1}
\end{equation}
where $x^{A}=x^{A}(x^{\lambda },x^{i})$, $g_{\mu \nu }=g_{\mu \nu
}(x^{\lambda })$ and $g_{ij}=g_{ij}(x^{k})$ are the metrics associated with
the $(1+3)$-dimensional world ($(1+3)$-world) and the $(3+1)$-dimensional
world ($(3+1)$-world), respectively. The fields $A_{\mu }^{i}=A_{\mu
}^{i}(x^{\alpha })$ and $B_{j}^{\mu }=B_{j}^{\mu }(x^{k})$ are a type of
gauge fields associated with the $(1+3)$-world and $(3+1)$-world
respectively.

Now, if we assume that the gauge fields $A_{\mu}^{i}$ and $B_{i}^{\mu}$ obey 
\begin{equation}
A_{\mu }^{i}B_{j}^{\mu }=0,  \label{kk2}
\end{equation}
and 
\begin{equation}
A_{\mu }^{i}B_{i}^{\nu }=0,  \label{kk3}
\end{equation}
then we find that the inverse $\gamma ^{AB}$ of the original metric $\gamma
_{AB}$ is given by 
\begin{equation}
\gamma ^{AB}(x^{\lambda },x^{i})=\left( 
\begin{array}{cc}
g^{\mu \nu }+B_{i}^{\mu }B_{j}^{\nu }g^{ij} & -g^{\mu \nu }A_{\nu
}^{i}-g^{ij}B_{j}^{\mu } \\ 
-g^{\nu \alpha }A_{\alpha }^{i}-g^{jj}B_{j}^{\nu } & g^{ij}+A_{\mu
}^{i}A_{\nu }^{j}g^{\mu \nu }%
\end{array}%
\right) .  \label{kk4}
\end{equation}%
Note that our assumptions (\ref{kk2}) and (\ref{kk3}) can be interpreted in
some sense that we are requiring that the gauge fields $A_{\mu }^{i}$ and $%
B_{i}^{\mu }$ are orthogonal.

With the help of (\ref{kk1}) it can be easily seen that the differential
line element in the $(4+4)$-space can be written as 
\begin{equation}
ds^{2}=(dx^{\mu }+B_{i}^{\mu }dx^{i})(dx^{\nu }+B_{j}^{\nu }dx^{j})g_{\mu
\nu } +(dx^{i}+A_{\mu }^{i}dx^{\mu })(dx^{j}+A_{\nu }^{j}dx^{\nu })g_{ij}.
\label{kk5}
\end{equation}
Thus, it follows that the basis 1-forms $\omega ^{\mu }$ and $\omega ^{i}$
read

\begin{equation}
\omega ^{\mu }=dx^{\mu }+B_{i}^{\mu }dx^{i},  \label{kk6}
\end{equation}
and 
\begin{equation}
\omega ^{i}=dx^{i}+A_{\nu }^{i}dx^{\nu }.  \label{kk7}
\end{equation}
Therefore, employing (\ref{kk6}) and (\ref{kk7}) the equation (\ref{kk5})
becomes

\begin{equation}
ds^{2}=\omega ^{\mu }\omega ^{\nu }g_{\mu \nu }+\omega ^{i}\omega ^{j}g_{ij}.
\label{kk8}
\end{equation}

Now, using (\ref{kk2}) and (\ref{kk3}), it is obtained from (\ref{kk6}) and (%
\ref{kk7}) that the dual basis can be written as

\begin{equation}
D_{\mu }=\partial _{\mu }-A_{\mu }^{i}\partial _{i}  \label{kk9}
\end{equation}
and 
\begin{equation}
D_{i}=\partial _{i}-B_{i}^{\mu }\partial _{\mu }.  \label{kk10}
\end{equation}
Of course, it is not difficult to see that the both derivatives $D_{\mu }$
and $D_{i}$, can be interpreted as covariant derivatives with connections $%
A_{\mu }^{i}$ and $B_{i}^{\mu }$, respectively.\newline

Let us now analize the anzats (\ref{kk1}) from another point of view. If we
write the metric $\gamma _{AB}$ in terms of a the vielbien field $%
E_{A}^{(C)} $ and the flat metric $\eta _{(CD)}=diag(-1,1,1,1,1,-1,-1,-1)$
in the form

\begin{equation}
\gamma _{AB}=E_{A}^{(C)}E_{B}^{(D)}\eta _{(CD)},  \label{kk11}
\end{equation}
then we can split (\ref{kk11}) as

\begin{equation}
\begin{array}{c}
\gamma _{\mu \nu }=e_{\mu }^{(\alpha )}e_{\nu }^{(\beta )}\eta _{(a\beta
)}+e_{\mu }^{(c)}e_{\nu }^{(d)}\eta _{(cd)} \\ 
\\ 
\gamma _{\mu i}=e_{\mu }^{(\alpha )}e_{i}^{(\beta )}\eta _{(a\beta )}+e_{\mu
}^{(c)}e_{i}^{(d)}\eta _{(cd)} \\ 
\\ 
\gamma _{ij}=e_{i}^{(\alpha )}e_{j}^{(\beta )}\eta _{(a\beta
)}+e_{i}^{(c)}e_{j}^{(d)}\eta _{(cd)}.%
\end{array}
\label{kk12}
\end{equation}

Thus, making the identifications $e_{\mu }^{(\alpha )}\equiv E_{\mu
}^{(\alpha )}$, $e_{i}^{(c)}\equiv E_{i}^{(c)}$, $g_{\mu \nu }=e_{\mu
}^{(\alpha )}e_{\nu }^{(\beta )}\eta _{(a\beta )}$, $%
g_{ij}=e_{i}^{(c)}e_{j}^{(d)}\eta _{(cd)}$, $A_{\mu }^{i}=e_{(c)}^{i}A_{\mu
}^{(c)}$ and $B_{i}^{\mu }=e_{(\alpha )}^{\mu }B_{i}^{(\alpha )}$ from (\ref%
{kk12}) we can obtain (\ref{kk1}).

Now, let us assume that the commutators of $e_{\mu }^{(\alpha )}$ and $%
e_{i}^{(c)}$ are given by

\begin{equation}
\lbrack e_{\mu }^{(\alpha )},e_{\mu }^{(\beta )}]=C_{(\gamma )}^{(\alpha
\beta )}e_{\mu }^{(\gamma )},  \label{kk13}
\end{equation}

\begin{equation}
\lbrack e_{\mu }^{(\alpha )},e_{i}^{(c)}]=0  \label{kk14}
\end{equation}

\begin{equation}
\lbrack e_{i}^{(c)},e_{i}^{(d)}]=C_{(e)}^{(cd)}e_{i}^{(e)},  \label{kk15}
\end{equation}
where $C_{(\gamma )}^{(\alpha \beta )}$ and $C_{(e)}^{(cd)}$ are structure
constants associated with the $(1+3)$-world and the $(3+1)$-world,
respectively. Thus, it is not difficult to see that (\ref{kk9}) and (\ref%
{kk10}) can now be written as

\begin{equation}
D_{\mu }=\partial _{\mu }-A_{\mu }^{(c)}e_{(c)}^{i}\partial _{i}
\label{kk16}
\end{equation}
and 
\begin{equation}
D_{i}=\partial _{i}-B_{i}^{(\alpha )}e_{(\alpha )}^{\mu }\partial _{\mu },
\label{kk17}
\end{equation}
being $e_{(\alpha )}^{\mu }$ and $e_{(c)}^{i}$ the inverse of $e_{\mu
}^{(\alpha )}$ and $e_{i}^{(c)}$, respectively. Therefore, using (\ref{kk13}%
)-(\ref{kk17}) it is straigthforward to see that the commutator of $D_{\mu } 
$ and $D_{i}$ leads to 
\begin{equation}
\lbrack D_{\mu },D_{\nu }]=-F_{\mu \nu }^{(c)}e_{(c)}^{i}\partial _{i},
\label{kk18}
\end{equation}

\begin{equation}
\lbrack D_{\mu },D_{i}]=0  \label{kk19}
\end{equation}

\begin{equation}
\lbrack D_{i},D_{j}]=-G_{ij}^{(\alpha )}e_{(\alpha )}^{\mu }\partial _{\mu },
\label{kk20}
\end{equation}
where

\begin{equation}
F_{\mu \nu }^{(c)}=\partial _{\mu }A_{\nu }^{(c)}-\partial _{\nu }A_{\mu
}^{(c)}+C_{(ed)}^{(c)}A_{\mu }^{(e)}A_{\nu }^{(d)}  \label{kk21}
\end{equation}
and 
\begin{equation}
G_{ij}^{(\alpha )}=\partial _{i}B_{j}^{(\alpha )}-\partial
_{j}B_{i}^{(\alpha )}+C_{(\gamma \sigma )}^{(a)}B_{i}^{(\gamma
)}B_{j}^{(\sigma )}  \label{kk22}
\end{equation}
are the corresponding field strengths.\newline

In a non-coordinate basis the metric $\gamma _{AB}\rightarrow \hat{\gamma}%
_{AB}$ becomes 
\begin{equation}
\hat{\gamma}_{AB}(x^{C})=\left( 
\begin{array}{cc}
g_{\mu \nu } & 0 \\ 
0 & g_{ij}%
\end{array}%
\right)  \label{kk23}
\end{equation}
while the connection $\Gamma _{ABC}$ results to be 
\begin{equation}
\Gamma _{ABC}=\frac{1}{2}(D_{C}\hat{\gamma}_{AB}+D_{B}\hat{\gamma}_{AC}-D_{A}%
\hat{\gamma}_{BC})+\frac{1}{2}(C_{ABC}+C_{ABC}-C_{BCA}),  \label{kk24}
\end{equation}
Here, $D_{A}$ is the covariant derivative associated with $\hat{\gamma}_{AB}$%
. In this manner, since $C_{\mu \nu \alpha }=0$ and $\hat{\gamma}_{\mu \nu
}=g_{\mu \nu }$ we find that

\begin{equation}
\Gamma _{\mu \nu \alpha }=\frac{1}{2}(g_{\mu \nu ,\alpha }+g_{\mu \alpha
,\nu }-g_{\nu \alpha ,\mu })=\{ \mu \nu \alpha \}.  \label{kk25}
\end{equation}
Similarly, because $C_{\mu \nu (\alpha )}=-F_{\mu \nu (c)}$, $C_{ij(\alpha
)}=-G_{ij(\alpha )}$ we arrive to the expressions 
\begin{equation}
\Gamma _{\mu \nu (c)}=-F_{\mu \nu (c)}  \label{kk26}
\end{equation}

\begin{equation}
\Gamma _{ij(\alpha )}=-G_{ij(\alpha )}.  \label{kk27}
\end{equation}
Finally, since $C_{ijk}=0$ and $\hat{\gamma}_{ij}=g_{ij}$ we obtain that

\begin{equation}
\Gamma _{ijk}=\frac{1}{2}(g_{ij,k}+g_{ik,j}-g_{jk,i})=\{ijk\}.  \label{kk28}
\end{equation}
Now, employing the formulae for the non-coordinate components of the Riemann
tensor 
\begin{equation}
\mathcal{R}_{BCD}^{A}=\partial _{C}\Gamma _{BD}^{A}-\partial _{D}\Gamma
_{BC}^{A}+\Gamma _{EC}^{A}\Gamma _{BD}^{E}-\Gamma _{ED}^{A}\Gamma
_{BC}^{E}-\Gamma _{BE}^{A}C_{CD}^{E},  \label{kk29}
\end{equation}
we can calculate the scalar curvature $\mathcal{R}=\gamma ^{BD}\mathcal{R}%
_{BAD}^{A}$. With this idea in mind, straightforward computations lead to

\begin{equation}
\mathcal{R}=\hat{R}+\tilde{R}+\frac{1}{2}F^{\mu \nu i}F_{\mu \nu }^{j}g_{ij}+%
\frac{1}{2}G^{ij\mu }G_{ij}^{\nu }g_{\mu \nu }.  \label{kk30}
\end{equation}

So, we can define the Einstein-Hilbert action in the form

\begin{equation}  \label{kk31}
S=\int dx^{(4+4)}\sqrt{\gamma }\, \mathcal{R}=S_{1}+S_{2},
\end{equation}
where the action corresponding for the $(1+3)$-space is 
\begin{equation}
S_{1}=a\int dx^{(1+3)}\sqrt{-\hat{g}}(\hat{R}+\frac{1}{2}F^{\mu \nu i}F_{\mu
\nu }^{j}\tilde{g}_{ij}^{0}) ,  \label{kk32}
\end{equation}
while for the $(3+1)$ we have

\begin{equation}
S_{2}=b\int dx^{(3+1)}\sqrt{-\tilde{g}}(\tilde{R}+\frac{1}{2}G^{ij\mu
}G_{ij}^{\nu }\hat{g}_{\mu \nu }^{0}).  \label{kk33}
\end{equation}
In here, $a$ and $b$ are volume constants and $\hat{g}$ and $\tilde{g}$
denote the determinant of $g_{\mu \nu }$ and $g_{ij}$, respectively. Besides 
$\hat{	R}$ is the scalar curvature associated with $g_{\mu \nu }$, while $%
\tilde{R} $ is the scalar curvature associated with $g_{ij}$. Moreover, we
defined the quantities $\tilde{g}_{ij}^{0}=\frac{1}{a}\int dx^{(3+1)}g_{ij}$
and $\hat{g}_{\mu \nu }^{0}=\frac{1}{b}\int dx^{(1+3)}g_{\mu \nu }$. An
interesting feature is that it seems that $S_{1}$ admits an intepretation of
a gravitational-Yang-Mills theory in a $(1+3)$-world, with $\tilde{g}%
_{ij}^{0}$ as a metric in a group manifold. Similarly, $S_{2}$ admits an
intepretation of a gravitational-Yang-Mills theory in a $(3+1)$-world, with $%
\tilde{g}_{ij}^{0}$ as a metric in a group manifold.\newline

Until this point, in order to motivate the subject we have focused in a $%
(4+4)$-world. However, our calculations are also valid for any $(n+n)$%
-world. Now, to derive an application of the formalism, we will study
solutions of the field equations static and with spherical symmetry. This
can be done in the simplest way; considering the gauge fields $B_{i}^{\mu }$
and $A_{\alpha }^{i}$ null. Hence, the line element (\ref{kk5}) acquires the
form 
\begin{equation}
ds_{(4+4)}^{2}=g_{\mu \nu }(x^{\alpha })dx^{\mu }dx^{\nu
}+g_{ij}(x^{k})dx^{i}dx^{j},  \label{ffe1}
\end{equation}%
where greek indices from now on run from $0$ to $3$ and represent the
coordinates of the $(1+3)$ space-time, whereas the capital latin indices run
from $4$ to $7$ and denote the coordinates of the $(3+1)$ space-time. Note
that both spaces are of complementary signature. Hence, the Levi-Civita
connection can be splitted up into 
\begin{equation}
^{(4+4)}\Gamma _{BD}^{A}=\, \left( 
\begin{array}{cc}
^{(1+3)}\Gamma _{\alpha \beta }^{\lambda } & 0 \\ 
0 & ^{(3+1)}\Gamma _{kl}^{i}%
\end{array}%
\right) .  \label{ffe2}
\end{equation}%
Consequently, according to the action (\ref{kk31}) the Einstein field
equations in vacuum lead to $^{(4+4)}\! \mathcal{R}_{AB}=0$ and consequently
one obtains%
\begin{equation}
\begin{array}{c}
R_{\mu \nu }=0, \\ 
\\ 
R_{ij}=0.%
\end{array}
\label{ffe3}
\end{equation}%
In what follows we shall show that this splitting no necessarily is a
trivial case.

\section{The Extended Kruskal-Szekeres Coordinates}

Now, in order to study spherically symmetric solutions of (\ref{ffe3}), we
introduce the $(4+4)$ line element having the spherical symmetry as

\begin{equation}
ds_{(4+4)}^{2}=-e^{2f(r)}dr^{2}+e^{2h(r)}dr^{2}+r^{2}d\Omega ^{2}+e^{2F(%
\tilde{r})}d\tilde{t}^{2}-e^{2H(\tilde{r})}d\tilde{r}^{2}-\tilde{r}^{2}d%
\tilde{\Omega}^{2},  \label{ffe4}
\end{equation}%
where $(t,r,\theta ,\phi )$ are the $(1+3)$-coordinates, $(\tilde{t},\tilde{r%
},\tilde{\theta},\tilde{\phi})$ are denoting the $(3+1)$-coordinates, $%
d\Omega ^{2}=d\theta ^{2}+sin^{2}\theta d\phi ^{2}$ and $d\tilde{\Omega}%
^{2}=d\tilde{\theta}^{2}+sin^{2}\tilde{\theta}d\tilde{\phi}^{2}$.\newline
It follows from (\ref{ffe3}) and (\ref{ffe4}) that the Schwarzschild metric
solution for a $(4+4)$-black hole reads 
\begin{equation}
ds^{2}=-\left( 1-\frac{r_{s}}{r}\right) dt^{2}+\frac{dr^{2}}{1-\frac{r_{s}}{r%
}}+r^{2}d\Omega ^{2}+\left( 1-\frac{r_{s}}{\tilde{r}}\right) d\tilde{t}^{2}-%
\frac{d\tilde{r}^{2}}{1-\frac{r_{s}}{\tilde{r}}}-\tilde{r}^{2}d\tilde{\Omega}%
^{2},  \label{ffe5}
\end{equation}%
where $r_{s}=2GM$ is the Schwarszchild radius, $M$ is associated with the
mass of the black hole and $(r,\theta ,\phi )$ are the spherical polar
coordinates ( In here we are using units such that the light velocity $c=1$%
). The solution corresponding to a (1+3)-signature then is given by 
\begin{equation}
ds^{2}=-\left( 1-\frac{r_{s}}{r}\right) dt^{2}+\frac{dr^{2}}{1-\frac{r_{s}}{r%
}}+r^{2}d\Omega ^{2}.  \label{1}
\end{equation}%
Now, we introduce the extended Kruskal--Szekeres coordinates in the form 
\begin{equation}
\begin{array}{c}
X=\epsilon \left[ \eta \left( \frac{r}{r_{s}}-1\right) \right] ^{1/2}e^{%
\frac{r}{2r_{s}}}cosh\left( \frac{t}{2R_{s}}\right) , \\ 
\\ 
T=\epsilon \left[ \eta \left( \frac{r}{r_{s}}-1\right) \right] ^{1/2}e^{%
\frac{r}{2r_{s}}}sinh\left( \frac{t}{2R_{s}}\right) ,%
\end{array}
\label{2}
\end{equation}%
where the new quantities $\epsilon $ and $\eta $ are parameters taking
values in the set \{$\pm 1$\}. Using the coordinate transformations (\ref{2}%
) one obtains the expression

\begin{equation}
X^{2}-T^{2}=\eta \left( \frac{r}{r_{s}}-1\right) e^{\frac{r}{r_{s}}}
\label{3}
\end{equation}%
and using the line element (\ref{1}) we arrive to 
\begin{equation}
\frac{4r_{s}^{3}}{r}e^{-\frac{r}{r_{s}}}\left( -dT^{2}+dX^{2}\right) =\eta %
\left[ -\left( 1-\frac{r_{s}}{r}\right) dt^{2}+\frac{dr^{2}}{1-\frac{r_{s}}{r%
}}\right] .  \label{4}
\end{equation}

Similarly, if we now write the extended Kruskal--Szekeres coordinates in the
alternative form

\begin{equation}
\begin{array}{c}
X=\epsilon \left[ \eta \left( \frac{r}{R_{s}}-1\right) \right] ^{1/2}e^{%
\frac{r}{2R_{s}}}sinh\left( \frac{t}{2R_{s}}\right) , \\ 
\\ 
T=\epsilon \left[ \eta \left( \frac{r}{R_{s}}-1\right) \right] ^{1/2}e^{%
\frac{r}{2R_{s}}}cosh\left( \frac{t}{2R_{s}}\right) ,%
\end{array}
\label{5}
\end{equation}%
it is not difficult to show that the analogous of (\ref{3}) reads

\begin{equation}
T^{2}-X^{2}=\eta \left( \frac{r}{r_{s}}-1\right) e^{\frac{r}{2r_{s}}}
\label{6}
\end{equation}%
and considering the line element (\ref{1}) we get the expression 
\begin{equation}
\frac{4r_{s}^{3}}{r}e^{-\frac{r}{r_{s}}}\left( -dT^{2}+dX^{2}\right) =\eta %
\left[ \left( 1-\frac{r_{s}}{r}\right) dt^{2}-\frac{dr^{2}}{1-\frac{r_{s}}{r}%
}\right] .  \label{7}
\end{equation}%
Notice that the equations (\ref{2}) and (\ref{5}) contain the both cases: $%
\epsilon =\pm 1$ and $\eta =\pm 1$. This means that we have $8$ different
ways of defining Kruskal--Szekeres coordinates, every case valid on
different regions of the space-time. We shall denote these cases as a
8-region approach.

Let us consider first the case $\eta =+1$. Then (\ref{2}) and (\ref{5})
leads to the Kruskal--Szekeres coordinates

\begin{equation}
\begin{array}{c}
X=\epsilon \left[ \left( \frac{r}{r_{s}}-1\right) \right] ^{1/2}e^{\frac{r}{%
2r_{s}}}cosh\left( \frac{t}{2R_{s}}\right) , \\ 
\\ 
T=\epsilon \left[ \left( \frac{r}{r_{s}}-1\right) \right] ^{1/2}e^{\frac{r}{%
2r_{s}}}sinh\left( \frac{t}{2R_{s}}\right) ,%
\end{array}
\label{8}
\end{equation}%
and%
\begin{equation}
\begin{array}{c}
X=\epsilon \left[ \left( \frac{r}{r_{s}}-1\right) \right] ^{1/2}e^{\frac{r}{%
2R_{s}}}sinh\left( \frac{t}{2R_{s}}\right) , \\ 
\\ 
T=\epsilon \left[ \left( \frac{r}{r_{s}}-1\right) \right] ^{1/2}e^{\frac{r}{%
2R_{s}}}cosh\left( \frac{t}{2R_{s}}\right) ,%
\end{array}
\label{9}
\end{equation}%
respectively. These transformations corresponds to values of $r$ such that $%
r>r_{s}$.

While if we consider $\eta =-1$ in (\ref{2}) and (\ref{5}) the corresponding
Kruskal-Szekeres coordinates are given by

\begin{equation}
\begin{array}{c}
X=\epsilon \left[ -\left( \frac{r}{r_{s}}-1\right) \right] ^{1/2}e^{\frac{r}{%
2r_{s}}}cosh\left( \frac{t}{2R_{s}}\right) , \\ 
\\ 
T=\epsilon \left[ -\left( \frac{r}{r_{s}}-1\right) \right] ^{1/2}e^{\frac{r}{%
2r_{s}}}sinh\left( \frac{t}{2R_{s}}\right) ,%
\end{array}
\label{10}
\end{equation}%
and%
\begin{equation}
\begin{array}{c}
X=\epsilon \left[ -\left( \frac{r}{r_{s}}-1\right) \right] ^{1/2}e^{\frac{r}{%
2R_{s}}}sinh\left( \frac{t}{2R_{s}}\right) , \\ 
\\ 
T=\epsilon \left[ -\left( \frac{r}{r_{s}}-1\right) \right] ^{1/2}e^{\frac{r}{%
2R_{s}}}cosh\left( \frac{t}{2R_{s}}\right) ,%
\end{array}
\label{11}
\end{equation}%
respectively. Now these transformations correspond to values of the radius $%
r $ such that $r_{s}<r$.

If we now regard $\epsilon =+1$ it is easy to see that (\ref{8})-(\ref{11})
yield

\begin{equation}
\begin{array}{c}
X=\left[ \left( \frac{r}{r_{s}}-1\right) \right] ^{1/2}e^{\frac{r}{2r_{s}}%
}cosh\left( \frac{t}{2R_{s}}\right) , \\ 
\\ 
T=\left[ \left( \frac{r}{r_{s}}-1\right) \right] ^{1/2}e^{\frac{r}{2r_{s}}%
}sinh\left( \frac{t}{2R_{s}}\right) ,%
\end{array}
\label{12}
\end{equation}%
\begin{equation}
\begin{array}{c}
X=\left[ \left( \frac{r}{r_{s}}-1\right) \right] ^{1/2}e^{\frac{r}{2R_{s}}%
}sinh\left( \frac{t}{2R_{s}}\right) , \\ 
\\ 
T=\left[ \left( \frac{r}{r_{s}}-1\right) \right] ^{1/2}e^{\frac{r}{2R_{s}}%
}cosh\left( \frac{t}{2R_{s}}\right) ,%
\end{array}
\label{13}
\end{equation}

\begin{equation}
\begin{array}{c}
X=\left[ -\left( \frac{r}{r_{s}}-1\right) \right] ^{1/2}e^{\frac{r}{2r_{s}}%
}cosh\left( \frac{t}{2R_{s}}\right) , \\ 
\\ 
T=\left[ -\left( \frac{r}{r_{s}}-1\right) \right] ^{1/2}e^{\frac{r}{2r_{s}}%
}sinh\left( \frac{t}{2R_{s}}\right) ,%
\end{array}
\label{14}
\end{equation}

\begin{equation}
\begin{array}{c}
X=\left[ -\left( \frac{r}{r_{s}}-1\right) \right] ^{1/2}e^{\frac{r}{2R_{s}}%
}sinh\left( \frac{t}{2R_{s}}\right) , \\ 
\\ 
T=\left[ -\left( \frac{r}{r_{s}}-1\right) \right] ^{1/2}e^{\frac{r}{2R_{s}}%
}cosh\left( \frac{t}{2R_{s}}\right) ,%
\end{array}
\label{15}
\end{equation}%
respectively. While if in (\ref{8})-(\ref{11}) we choose $\epsilon =-1$ we
obtain

\begin{equation}
\begin{array}{c}
X=-\left[ \left( \frac{r}{r_{s}}-1\right) \right] ^{1/2}e^{\frac{r}{2r_{s}}%
}cosh\left( \frac{t}{2R_{s}}\right) , \\ 
\\ 
T=-\left[ \left( \frac{r}{r_{s}}-1\right) \right] ^{1/2}e^{\frac{r}{2r_{s}}%
}sinh\left( \frac{t}{2R_{s}}\right) ,%
\end{array}
\label{16}
\end{equation}%
\begin{equation}
\begin{array}{c}
X=-\left[ \left( \frac{r}{r_{s}}-1\right) \right] ^{1/2}e^{\frac{r}{2R_{s}}%
}sinh\left( \frac{t}{2R_{s}}\right) , \\ 
\\ 
T=-\left[ \left( \frac{r}{r_{s}}-1\right) \right] ^{1/2}e^{\frac{r}{2R_{s}}%
}cosh\left( \frac{t}{2R_{s}}\right) ,%
\end{array}
\label{17}
\end{equation}

\begin{equation}
\begin{array}{c}
X=-\left[ -\left( \frac{r}{r_{s}}-1\right) \right] ^{1/2}e^{\frac{r}{2r_{s}}%
}cosh\left( \frac{t}{2R_{s}}\right) , \\ 
\\ 
T=-\left[ -\left( \frac{r}{r_{s}}-1\right) \right] ^{1/2}e^{\frac{r}{2r_{s}}%
}sinh\left( \frac{t}{2R_{s}}\right) ,%
\end{array}
\label{18}
\end{equation}

\begin{equation}
\begin{array}{c}
X=-\left[ -\left( \frac{r}{r_{s}}-1\right) \right] ^{1/2}e^{\frac{r}{2R_{s}}%
}sinh\left( \frac{t}{2R_{s}}\right) , \\ 
\\ 
T=-\left[ -\left( \frac{r}{r_{s}}-1\right) \right] ^{1/2}e^{\frac{r}{2R_{s}}%
}cosh\left( \frac{t}{2R_{s}}\right) .%
\end{array}
\label{19}
\end{equation}

Considering all eight transformations (\ref{12})-(\ref{19}) an $8$-region
spacetime structure is obtained. In the four expressions (\ref{12}), (\ref%
{15}), (\ref{16}) and (\ref{19}) we recognize the traditionally $4$-region
spacetime described for instance in the Ref. \cite{03}. What seems to be new
are the transformations (\ref{13}),(\ref{14}), (\ref{17}) and (\ref{18}).
These give another $4$-region. So if one add the traditional and the new $4$%
-region one gets a $8$-region space-time structure. However, one may ask:
What are the key result that distinguish these two $4$-regions?. In the
following section we shall show that the traditional $4$-region can be
associated with a space-time of $(1+3)$-signature, while the new $4$-region
must be associated with a $(3+1)$-signature. Our final conclusion will be
that the $8$-region Kruskal--Szekeres transformation corresponds to a world
with $(4+4)$-signature.

\section{(4+4)-space-time structure and mirror black and white holes}

In order to obtain information about the physical and geometrical meaning of
the new 4-region we proceed as follows.\newline

First, it is important to note that in all the process of the
Kruskal--Szekeres transformations the angular part of (\ref{1}), which is
given by

\begin{equation}
ds_{(2)}^{2}\equiv r^{2}(d\theta ^{2}+sin^{2}\theta d\phi ^{2})  \label{20}
\end{equation}%
has not been considered. In fact, the traditional Kruskal--Szekeres method
focus only in the first part of (\ref{1}), namely

\begin{equation}
ds_{(1)}^{2}\equiv -(1-\frac{r_{s}}{r})dt^{2}+\frac{dr^{2}}{1-\frac{r_{s}}{r}%
}.  \label{21}
\end{equation}
Of course, it is not difficult to see that when $r\rightarrow \infty $, the
equations (\ref{20}) and (\ref{21}) lead to a flat world of $(1+3)$%
-signature. It turns out that the $4$-region transformations (\ref{12}), (%
\ref{15}), (\ref{16}) and (\ref{19}) are compatible with the $(1+3)$%
-signature when we set $\eta =+1$ in (\ref{4}) and $\eta =-1$ in (\ref{7}).

On the other hand, the angular part for the (3+1)-part of the metric (\ref%
{ffe5}) is given by

\begin{equation}
dl_{(2)}^{2}\equiv -r^{2}(d\theta ^{2}+sin^{2}\theta d\phi ^{2}),  \label{22}
\end{equation}%
while its radial part reads

\begin{equation}
dl_{(1)}^{2}\equiv -\left[-\left(1-\frac{r_{s}}{r}\right)dt^{2}+\frac{dr^{2}%
}{1-\frac{r_{s}}{r}}\right].  \label{23}
\end{equation}%
Thus, it seems evident from (\ref{22}) and (\ref{23}) that when $%
r\rightarrow \infty $ a flat world of $(3+1)$-signature is obtained.
Something remarkable is that the transformations (\ref{13}),(\ref{14}), (\ref%
{17}) and (\ref{18}) can be made compatible with (\ref{22}) and (\ref{23})
if we substitute $\eta =-1$ in (\ref{4}) and $\eta =+1$ in the expression (%
\ref{7}). Hence, in order to distinguish the black-hole solution (\ref{22})
and (\ref{23}) from the usual one we shall call it mirror-black-hole. This
name can be justified because while ordinary black-holes live in $(1+3)$%
-dimensions the mirror-black-hole lives in a mirror world of $(3+1)$%
-dimensions.

Let us now analyze, from the perspective of Kruskal--Szekeres coordinates,
geometrical implications when $r=0$ in both cases: black and mirror-black
holes. Observe first that when $r=0$ the equations (\ref{3}) and (\ref{6})
lead to

\begin{equation}
X^{2}-T^{2}=-\eta ,  \label{24}
\end{equation}
and

\begin{equation}
T^{2}-X^{2}=\eta ,  \label{25}
\end{equation}%
respectively. In the traditional steps of Kruskal--Szekeres coordinates the
transformations (\ref{13}),(\ref{14}), (\ref{17}) and (\ref{18}) are
considered in such a way that in (\ref{24}) and (\ref{25}) the value $\eta
=+1$ is taken. It is evident that in this case the two expressions (\ref{24}%
) and (\ref{25}) lead to exactly the same equation, namely

\begin{equation}
T^{2}-X^{2}=+1.  \label{26}
\end{equation}%
Of course, this expression corresponds to an hyperbola in the plane $X$ and $%
T$. The branch of this hyperbola in the positive values of $T$ is identified
with the true singularity of the black-hole. The other branch of the
hyperbola for negative values of $T$ is associated with the singularity of a
possible white hole. On the other hand if in the equations (\ref{24}) and (%
\ref{25}) we use $\eta =-1$ and considering the transformations (\ref{13}), (%
\ref{14}), (\ref{17}) and (\ref{18}), we obtain that instead of (\ref{26})
we have now the formula

\begin{equation}
T^{2}-X^{2}=-1.  \label{27}
\end{equation}%
It is easy to verify that hyperbola in this case will correspond to positive
and negative values of $X$. This means that we shall have not only a
singularity associated with the mirror-black-hole but also to a kind of
mirror-white-hole. Hence we have mirror black and white holes in a $(3+1)$%
-signature space-time, which from a global point of view we can interpret
that in order to describe black/white and mirror black/white holes in a
single gravitational setting, a theory with a (4+4)-space-time structure
would be necessary.

\section{Linearized gravity in (4+4)-dimensions.}

Another non-trivial aspect of general relativity in $(4+4)$-dimensions
emerges from linearized gravity. It turns out that in the year 2003, Nishino
and Rajpoot [40] presented a self-dual N = (1,0) supergravity in Euclidean $%
8 $-dimensions. Their method consisted in considering the self-duality
concept of the curvature in terms of a four-form octonion structure
constants $\eta ^{ABCD}$. This motivates to consider the Rarita-Schwinger
field equation in eight dimensions

\begin{equation}
\eta ^{ABCD}\gamma _{B}\partial _{C}\Psi _{D}=0,  \label{N1}
\end{equation}%
in completely analogy to the case of four dimensions

\begin{equation}
\epsilon ^{\mu \nu \sigma \alpha }\gamma _{\nu }\partial _{\sigma }\Psi
_{\alpha }=0.  \label{N2}
\end{equation}

On the other hand, a classical description of (67)

\begin{equation}
S_{\mu }^{\alpha }=\epsilon _{\mu }^{~\tau \sigma \alpha }\theta _{\tau
}p_{\sigma }=0.  \label{N3}
\end{equation}
in terms of anticommuting Grassmann variables $\theta _{\tau }$ leads to the
remarkable result that $S_{\mu }^{\alpha }$ can be undertood as the square
root of linearized gravity, in the sense that

\begin{equation}
\left \{ S_{\mu }^{\alpha },S_{\nu }^{\beta }\right \} =H_{\mu \nu }^{\alpha
\beta },  \label{N4}
\end{equation}%
where

\begin{equation}
\begin{array}{c}
H_{\mu \nu }^{\alpha \beta }=-\delta _{\mu }^{\alpha }\delta _{\nu }^{\beta
}p^{\lambda }p_{\lambda }-\eta ^{\alpha \beta }p_{\mu }p_{\nu }+\delta _{\mu
}^{\alpha }p_{\nu }p^{\beta }+\delta _{\mu }^{\alpha }p_{\nu }p^{\beta } \\ 
\\ 
-\eta _{\mu \nu }p^{\alpha }p^{\beta }+\eta _{\mu \nu }\eta ^{\alpha \beta
}p^{\lambda }p_{\lambda }.%
\end{array}
\label{N5}
\end{equation}%
corresponds to first class constraint associated with linearized gravity
(see Ref. [41] for details). Of course, this result is inspired in the
supersymmetric $\frac{1}{2}$-spin formalism. In this case the Dirac equation

\begin{equation}
(\gamma ^{\mu }\hat{p}_{\mu }+m)\psi =0  \label{N6}
\end{equation}%
is considered as a first class constraint

\begin{equation}
S=\theta ^{\mu }p_{\mu }+m\approx 0  \label{N7}
\end{equation}%
and in this way one proves the relation

\begin{equation}
\{S,S\}=H,  \label{N8}
\end{equation}%
meaning that $S$ is the square root of the constraint

\begin{equation}
H=p^{\mu }p_{\mu }+m^{2}\approx 0.  \label{N9}
\end{equation}%
Of course, at the quantum level this constraint can be associated with the
Klein-Gordon equation

\begin{equation}
(\hat{p}^{\mu }\hat{p}_{\mu }+m^{2})\varphi =0.  \label{N10}
\end{equation}%
One of our motivation is to see whether one can follow similar steps to the
case of the gravitino field equation (66). For this purpose in analogy to
(68) let us associate to (66) the constraint

\begin{equation}
\mathcal{S}_{A}^{E}=\eta _{A}^{~EDF}\theta _{D}p_{F}\approx 0.  \label{N11}
\end{equation}

Our goal is to determine

\begin{equation}
\{ \mathcal{S}_{A}^{E},\mathcal{S}_{B}^{F}\}=\mathcal{H}_{AB}^{EF}.
\label{N12}
\end{equation}%
First one observe that 
\begin{equation}
\mathcal{H}_{AB}^{EF}=\eta _{A}^{~GHE}p_{H}\eta _{BG}^{~~QF}p_{Q}.
\label{N13}
\end{equation}%
This expression can be rewritten as

\begin{equation}
\mathcal{H}_{AB}^{EF}=\eta _{AR}\eta ^{FP}\eta ^{ERHG}\eta _{PBQG}p_{H}p^{Q}.
\label{N14}
\end{equation}

Now, it is known that the octonionic structure constants $\eta ^{ABCD}$
satifies [42]-[43],

\begin{equation}
\eta ^{ABCD}\eta _{EFGD}=\delta _{EFG}^{ABC}+\Sigma _{EFG}^{ABC},
\label{N15}
\end{equation}%
where $\Sigma _{EFG}^{ABC}$ is given by

\begin{equation}
\begin{array}{cc}
\Sigma _{EFG}^{ABC}= & \eta _{EF}^{AB}\delta _{G}^{C}+\eta _{EF}^{BC}\delta
_{G}^{A}+\eta _{EF}^{CA}\delta _{G}^{B} \\ 
&  \\ 
& \eta _{FG}^{AB}\delta _{E}^{C}+\eta _{FG}^{BC}\delta _{R}^{A}+\eta
_{FG}^{CA}\delta _{E}^{B} \\ 
&  \\ 
& \eta _{GE}^{AB}\delta _{F}^{C}+\eta _{GE}^{BC}\delta _{F}^{A}+\eta
_{GE}^{CA}\delta _{F}^{B}.%
\end{array}
\label{N16}
\end{equation}

Hence, substituting (80) and (81) into (79) yields

\begin{equation}
\mathcal{H}_{AB}^{EF}=H_{AB}^{EF}+\Omega _{AB}^{EF}.  \label{N17}
\end{equation}%
Here $H_{AB}^{EF}$ has exatly the same form than (70) but in eight
dimensions and $\Omega _{AB}^{EF}$ is given by

\begin{equation}
\Omega _{AB}^{EF}=\eta _{AQ}\eta ^{FR}\Sigma _{RBH}^{EQG}p_{G}p^{H},
\label{N18}
\end{equation}%
which in virtue of (81) leads to

\begin{equation}
\begin{array}{cc}
\Omega _{AB}^{EF}= & -\eta _{~~~~AB}^{EF}p_{Q}p^{Q}-\eta
_{~ABD}^{F}p^{D}p^{E}+\eta _{~~~D}^{EF}p^{D}p_{A} \\ 
&  \\ 
& +\eta _{~ABD}^{E}p^{D}p^{F}-\eta _{~~~AD}^{EF}p^{D}p_{B}.%
\end{array}
\label{N19}
\end{equation}%
Observe that 
\begin{equation}
\Omega _{AB}^{EF}=-\Omega _{AB}^{FE}=-\Omega _{BA}^{EF}.  \label{N20}
\end{equation}

Thus, since $H_{AB}^{EF}$ is obtained from the Riemann tensor we see that
the result (85) implies a non-trivial additional term $\Omega _{AB}^{EF}$ in
the Einstein field equations.

\section{Final Remarks}

In this work we have introduced a new type of $(4+4)$-Kaluza-Klein theory of
gravity. As an application of the theory we have studied spherically
symmetric solutions in va\-cuum, in the particular case where the gauge
fields $A_{\alpha}^{i}$ and $B_{i}^{\mu}$ are null. Hence, considering
generalized Kruskal-Szekeres coordinate transformations, we obtain two
4-regions represen\-ting black/white hole solutions in a (1+3)-space-time
signature and mirror black/white hole solutions in a (3+1)-space-time
signature, respectively.

The present developments clearly point towards a more general theory which
combines black/white-holes and mirror-balck/white-holes. Since the
corresponding signatures are $(1+3)$ and $(3+1)$ the natural choice for such
a generalized theory seems to be a framework in $(4+4)$-dimensions. This is
in part due to the fact that the interesting splitting $(4+4)=(1+3)+(3+1)$
is valid. Surprisingly, the $(4+4)$-dimensional theory can be understood as
a particular case of the so called double field theories \cite{19}. It may
be interesting for further research to explore this possible connection.

Moreover, in Ref. [20] it was shown that the Dirac equation in $(4+4)$%
-dimensions leads to the surprising result that a complex spinor associated
with $\frac{1}{2}$-spinor in $(1+3)$-dimensions can be understood as a
Majorana-Weyl spinor in $(4+4)$-dimensions. So, one would expect that the
Majorana-Weyl vector-spinor $\Psi _{D}$ of the Rarita-Schwinger field
equation in $(4+4)$-dimensions (66) can be associated with a complex spinor
vector-spinor in $(1+3)$-dimensions. Furthermore, using Cayley
hyperdeterminant it was shown that there exist black-hole/qubit
correspondence in (4+4)-dimensions [44], which in turn can be linked to
oriented matroid theory (see Refs [45]-[48] and references therein).
Finally, it is worth mentioning that an Ashtekar formalism in eight
dimensions has been developed in Ref. [49]. Therefore, one may expect for a
further work may emerge when the present formalism of general relativity in $%
(4+4)$-dimensions becomes connected with such fascinating developments.

\section*{Acknowledgements}

\noindent J. A. Nieto acknowledges Centro Universitario de Ciencias Exactas
e Ingenierias and Centro Universitario de los Valles of Universidad de
Guadalajara for hospitality and financial support. J.E.Madriz-Aguilar
acknowledges CONACYT M\'{e}xico and Centro Universitario de Ciencias Exactas
e Ingenierias of Universidad de Guadalajara for financial support.


\bigskip

\begin{thebibliography}{99}
\bibitem{k1} H. Nicolai and C. Wetterich, Phys. Lett. B \textbf{150} (1985)
347.

\bibitem{k2} C. Wetterich, Nucl. Phys. B \textbf{255} (1985) 480.

\bibitem{k3} S. Randjbar-Daemi and C. Wetterich, Phys. Lett. B \textbf{166}
(1986) 65.

\bibitem{k4} T. Liko, Phys. Lett. B \textbf{617} (2005) 193; e-Print:
hep-th/0505049.

\bibitem{k5} J. E. Madriz Aguilar and M. Bellini, Phys. Lett. B \textbf{619}
(2005) 208; e-Print: gr-qc/0503045.

\bibitem{k6} A. Raya Montano, J. E. Madriz Aguilar and and M. Bellini, Adv.
Stud. Theor. Phys. \textbf{1} (2007) 281;

\bibitem{k7} J. E. Madriz Aguilar, Mauricio Bellini and F. Astorga Saenz;
Published in In *Reimer, A. (ed.): Quantum cosmology research trends* 93-120
, e-Print: gr-qc/0411074.

\bibitem{k8} M. Bellini, Phys. Lett. B \textbf{609} (2005) 187;
gr-qc/0410143.

\bibitem{k9} J. E. Madriz Aguilar and M. Bellini, Eur. Phys. J. C \textbf{42}
(2005) 349; gr-qc/0408054.

\bibitem{k10} J. E. Madriz Aguilar and M. Bellini Eur. Phys. J. C \textbf{38}
(2004) 367;\ hep-th/0406268.

\bibitem{k11} J. E. Madriz Aguilar and Mauricio Bellini, Phys. Lett. B%
\textbf{\ 596} (2004) 116; gr-qc/0405024.

\bibitem{k12} J. E. Madriz Aguilar and Mauricio Bellini, Eur. Phys. J. C 
\textbf{38} (2004) 123; gr-qc/0402108.

\bibitem{k13} J. Ponce de Leon, Int. J. Mod. Phys. D \textbf{12 }(2003) 757;
gr-qc/0209013

\bibitem{k14} J. Ponce de Leon, Gen. Rel. Grav. \textbf{35 }(2003) 1365.
gr-qc/0207108.

\bibitem{k15} Y. M. Cho, J. Math. Phys. \textbf{16} (1975) 2029.

\bibitem{k16} Y. M. Cho and P. G. O. Freund, Phys. Rev. D \textbf{12 }%
(1975)1711.

\bibitem{k17} T. Fukuyama, Gen. Rel. Grav. \textbf{14} (1982) 729.

\bibitem{k18} J. H. Yoon, Class. Quant. Grav. \textbf{16} (1999) 1863;
gr-qc/0003060.

\bibitem{k19} J. H. Yoon, Phys. Lett. B \textbf{451} (1999) 296.

\bibitem{k20} J. A. Nieto and M. Espinoza, Int. J. Geom. Meth. Mod. Phys. 
\textbf{14} (2016) 1750014.

\bibitem{01} M. D. Kruskal, Phys. Rev. \textbf{119} (1960) 1743.

\bibitem{02} G. Szekeres, Publ. Math. Debrecen \textbf{7} (1960) 285.

\bibitem{03} C. M. Misner, K. S. Thorne, J. A. Wheeler, \textit{Gravitation}%
, (Ed. W. H. Freeman and Company, 1973).

\bibitem{04} S. M. Carroll, \textit{Spacetime and Geometry}, (Addison
Wesley, 2004).

\bibitem{05} M. J. Duff, J. Kalkkinen, Nucl. Phys. B \textbf{760} (2007) 64;
hep-th/0605274.

\bibitem{06} M.J. Duff, J. Kalkkinen, Nucl.Phys. B \textbf{758} (2006) 161;
hep-th/0605273.

\bibitem{07} C. Chevalley, The Algebraic Theory of Spinors (Columbia
University Press, 1954).

\bibitem{08} S. Kobayashi, K. Nomizu, "Foundations of Differential
Geometry", Vols. 1 and 2 (New ed.), Wiley-Interscience, ISBN 0-471-15733-3
(1996). (2007).

\bibitem{09} Y. Choquet-Bruhat and C. DeWitt-Morette, Analysis, Manifols and
Physics. Part II, Applications (North-Holland, Amsterdam, 1989).

\bibitem{10} P. Deligne, Notes on Spinors, in : Quantum Fields and Strings:
A Course for Mathematicians (American Mathematical Society, 1999).

\bibitem{11} R. D'Auria, S. Ferrara, M. A. Lledo and V. S. Varadarajan,
Spinor alge- bras, J. Geom. Phys. \textbf{40} (2001) 101.

\bibitem{12} P. G. O. Freund, Introduction to Supersymmetry (Cambridge
University Press, 1988).

\bibitem{13} C. Hull, JHEP \textbf{9811} (1998) 017; hep-th/9807127.

\bibitem{14} C. Hull, JHEP \textbf{9807} (1998) 021; hep-th/9806146.

\bibitem{15} C. Hull and R. R. Khuri, Nucl. Phys. B\textbf{536} (1998) 219;
hep-th/9808069.

\bibitem{16} M. A. De Andrade, M. Rojas and F. Toppan, Int. J. Mod. Phys. A 
\textbf{16} (2001) 4453; hep-th/0005035.

\bibitem{17} M. Rojas, M. A. De Andrade, L. P. Colatto, J. L. Matheus-Valle,
L. P. G. De Assis and J. A. Helayel-Neto, ``Mass Generation and Related
Issues from Exotic Higher Dimensions''; hep-th/1111.2261.

\bibitem{18} M. A. De Andrade and I.V. Vancea., ``Action for spinor fields
in arbitrary dimensions''; hep-th/0105025.

\bibitem{19} C. Hull, B. Zwiebach, JHEP \textbf{0909} (2009) 099;
hep-th/0904.4664.

\bibitem{20} H. Nishino and S. Rajpoot, Phys. Lett. B \textbf{564} (2003)
269-279; hep-th/0302059

\bibitem{21} J. A. Nieto, O. Obregon, Phys. Lett. A \textbf{175} (1993)
11-13.

\bibitem{22} A. R. Dundarer, F. Gursey and C. H. Tze, J. Math. Phys. \textbf{%
25} (1984) 1496.

\bibitem{23} A. R. Dundarer and F. Gursey, J. Math. Phys. \textbf{32} (1991)
1178.

\bibitem{24} J. A. Nieto, Phys. Lett. B \textbf{718} (2013) 1543;
arXiv:1210.0928 [hep-th].

\bibitem{25} J. A. Nieto, Nucl. Phys. B \textbf{883} (2014) 350;
arXiv:1402.6998 [hep-th].

\bibitem{26} J. A. Nieto, Phys. Lett. B \textbf{692} (2010) 43;
arXiv:1004.5372 [hep-th].

\bibitem{27} J. A. Nieto, Adv.Theor. Math. Phys. \textbf{8} (2004) no.1,
177; hep-th/0310071.

\bibitem{28} J. A. Nieto, Adv. Theor. Math. Phys. \textbf{10} (2006) no.5,
747; hep-th/0506106.

\bibitem{29} J. A. Nieto, Class. Quant. Grav. \textbf{22} (2005) 947-955;
hep-th/0410260
\end{thebibliography}
\end{document}